\documentclass[aps,prl,twocolumn,groupedaddress]{revtex4-1}

\usepackage{here}
\usepackage{amsmath}
\usepackage{graphicx}

\def\apj{ApJ}

\def\apjs{ApJS}

\def\prd{Phys.\ Rev.\ D}

\begin{document}


\title{Circular polarizations of gravitational waves from core-collapse  supernovae: \\a clear indication of rapid rotation}


\author{Kazuhiro Hayama$^{1,2}$, Takami Kuroda$^{3}$, Ko Nakamura$^{4}$, Shoichi Yamada$^{4,5}$}
\affiliation{$^1$KAGRA Observatory, Institute for Cosmic Ray Research, University of Tokyo, Japan }
\affiliation{$^2$Gravitational Wave Project Office, National Astronomical Observatory of Japan, Japan}
\affiliation{$^3$Department of Physics, University of Basel, Switzerland}
\affiliation{$^4$Advanced Research Institute for Science and Engineering, Waseda University}
\affiliation{$^5$Science \& Engineering, Waseda University, Japan}


\date{\today}

\begin{abstract}
We propose to employ the circular polarization of gravitational waves emitted 
by core-collapse supernovae as an unequivocal indication of rapid rotation deep in their cores.
It has been demonstrated by three dimensional simulations that non-axisymmetric accretion 
flows may develop spontaneously via hydrodynamical instabilities in the post-bounce
cores. It is not surprising then that the gravitational waves emitted by such fluid motions are
circularly polarized. We show in this letter that a network of the second generation detectors
of gravitational waves worldwide may be able to detect such polarizations up to the opposite side
of Galaxy as long as the rotation period is shorter than a few seconds prior to collapse.

\end{abstract}


\maketitle


\textbf{Introduction{\rm---}}
A direct detection of a gravitational wave (GW) is not a matter of if but a matter of when. Three km-scale 
laser interferometers, LIGO Hanford (H), LIGO Livingston (L) and Virgo (V), have been in operation at the designed sensitivities for years and an upgrade to the second generation will be completed in a couple of years~\cite{aligoweb,advirgoweb}. KAGRA (K), 
another second-generation GW detector in Japan will soon join them~\cite{kagraweb}. The network of these advanced 
GW detectors will be then poised to observe various targets in the universe, in which core-collapse 
supernovae are included. It is hence an urgent task for researchers to get ready for the first detection.

The core-collapse supernova (CCSN) is an energetic explosion of massive stars at the end of their lives. 
Gravitational collapse of the central core precedes the expansion of the envelope as we observe it
optically. How the initial implosion of the core leads eventually to the explosion of the star has been
an unsolved problem for decades~\cite{Janka2012}. The current scenario goes as follows: the collapse proceeds 
until the central density reaches the nuclear saturation density, at which point nuclear forces decelerate 
the contraction of the inner part of the core; a shock wave is then produced by core bounce and 
propagates outward; the shock is not strong enough initially to expel infalling matter and 
stagnates inside the core; it is then somehow revived to propagate through the entire envelope 
of the star and produce an explosion when it reaches the stellar surface. For the moment, neutrinos 
emitted copiously from a proto-neutron star are the most promising agents to reinvigorate the stalled shock 
wave.

Whatever its mechanism, it will not be a surprise that CCSN, in which a solar-mass-scale gas moves 
dynamically on a time scale of milliseconds, is an important target for the advanced GW detectors~\cite{ott2009,kotake2013}. 
Core bounce, for example, is one of the most violent events in CCSN and its GW emissions have 
been studied extensively over the years. As the neutrino-heating mechanism is scrutinized, on the other 
hand, it becomes recognized that the turbulence induced by hydrodynamical instabilities in the post-shock 
accretion flows can be another source of GW. In fact, multi-dimensional simulations of CCSN commonly
observe the so-called standing accretion shock instability (SASI) and convection grow from tiny seed 
fluctuations and render the post-shock flow highly turbulent. When such fully-fledged turbulent flows hit
the proto-neutron star surface, stochastic GWs are produced at levels observable at Galactic distances~\cite{ott2009,kotake2013,mueller2013}.
Since the Galactic supernova will be a once-in-a-life event, it is critically important for us to be able to 
extract as much information as possible from it when it really occurs.
Stars are rapid rotators in general (For example, see the observations of the main sequence~\cite{agudelo2015}). It is highly unknown, however, how fast they are rotating deep inside them just prior to collapse. There is a ballpark of theoretical estimates at present: short periods of a few seconds are obtained if
no magnetic braking is taken into account whereas more than ten times longer periods are common
outcomes if the angular momentum transfer via magnetic fields is assumed~\cite{maeder2012}. Rotation is hence
a major uncertainty in the stellar structure and evolution theory, one of the foundations of astrophysics. GWs may provide us with a rare chance to reveal it.

It turns out that it is not so easy, though. As a matter of fact, there is a history in the research of stellar rotations with GW. Hayama et al.~\cite{hayama08}, for example, proposed a method to employ the sign of the second largest peak in the GW signal from core bounce. The methodology was criticized later, however, by Abdikamalov et al.~\cite{abdikamalov2014}, who conducted a larger number of numerical simulations and proposed instead another method. It was still based on the theoretical catalogue of GW wave forms obtained in their simulations of various supernovae, which unfortunately may not be correct. 


In this letter we propose a simpler and clearer method to probe a rapid rotation in the core with GW signals alone once a detection of GW has been done on a network of detectors during their operations with stable outputs of Gaussian noise. In contrast to the previous studies~\cite{abdikamalov2014}, which focused on the waveforms of GW, particularly the amplitudes and characteristic frequencies, we pay attention here to circular polarizations 
of the GW emitted after core bounce. According to some recent three dimensional simulations, collapse 
of rapidly rotating cores of massive stars will lead to the formation of non-axisymmetric, spiral patterns in 
the post-shock accretion flows~\cite{kuroda2014,nakamura2014}. From an analogy to the GW from binaries, it is expected that the GW 
emitted in the direction of rotation axis will have circular polarizations at the frequency just twice 
the rotational frequency. Although this seems to be rather obvious, the issue is, of course, whether such polarizations, if any, are observable on the terrestrial GW detectors or not. In the following, 
we demonstrate that the answer is positive indeed for Galactic supernovae if rotation is rapid enough. Employing the noise-free GW waveforms calculated theoretically in one of the latest 3D simulations of rapidly rotating CCSNe~\cite{kuroda2014}, we reconstruct them for the 2nd generation GW detectors mentioned above, taking into account their noises appropriately. We then evaluate the Stokes parameters based on those reconstructed waveforms. It is stressed that this study would not have been possible without the right combination of (1) an emergence of non-axisymmetric structures that are sustained for many cycles of rotation, (2) appropriate mass, radius and rotation period of these structures and (3) more than two GW detectors with high sensitivities.

\textbf{Circular polarization of GW{\rm---}}
The polarization of GW is most conveniently described by the Stokes parameters\cite{2007PhRvL..99l1101S}, which are related with the ensemble averages of the GW amplitudes as follows:
\begin{equation}
\begin{aligned}
&\left (
\begin{array}{cc}
 \langle h_R(f,{\hat n})h_R(f',{\hat n}')^*\rangle & \langle h_L(f,{\hat n})h_R(f',{\hat n}')^*\rangle \\
 \langle h_R(f,{\hat n})h_L(f',{\hat n}')^*\rangle & \langle h_L(f,{\hat n})h_L(f',{\hat n}')^*\rangle
\end{array}
\right )\\
& = \frac{1}{4\pi} \delta^2_D({\hat n}-{\hat n}')\delta_D(f-f')\\
&\left (
\begin{array}{cc}
 I(f,{\hat n})+V(f,{\hat n}) & Q(f,{\hat n})-iU(f,{\hat n}) \\
 Q(f,{\hat n})+iU(f',{\hat n}) & I(f,{\hat n})-V(f,{\hat n})
\end{array}
\right ), &
\end{aligned}
\end{equation}
where $f$ is a frequency, ${\hat n}$ is a unit vector in the direction of the propagation, $\delta_D$ is the Dirac's delta function, and the amplitudes of the left\mbox{-}handed and right\mbox{-}handed modes of GW are denoted by $h_R:=(h_+-\mathrm{i}h_{\times})/\sqrt{2}$ and $h_L:=(h_++\mathrm{i}h_{\times})/\sqrt{2},$ respectively. Stokes parameter $V$ characterizes the asymmetry between the right\mbox{-} and left\mbox{-}handed modes whereas parameter $I(\geq|V|)$ corresponds to the total amplitudes. $\langle h_R h_L^*\rangle$ and $\langle h_L h_R^*\rangle$ represent the linear polarization modes which are proportional to $Q \pm \mathrm{i} U.$ In this letter we focus on the $V$ parameter. We reconstruct gravitational waveforms employing the coherent network analysis for the global network of the second-generation GW detectors~\cite{Klimenko:2005,hayama2007,2010NJPh...12e3034S,hayama15}. The V\mbox{-}parameter is then calculated from reconstructed waveforms, $h_+$, $h_\times$.

\begin{figure}[t]
\begin{center}
\includegraphics[width=0.9\linewidth,height=6.4cm]{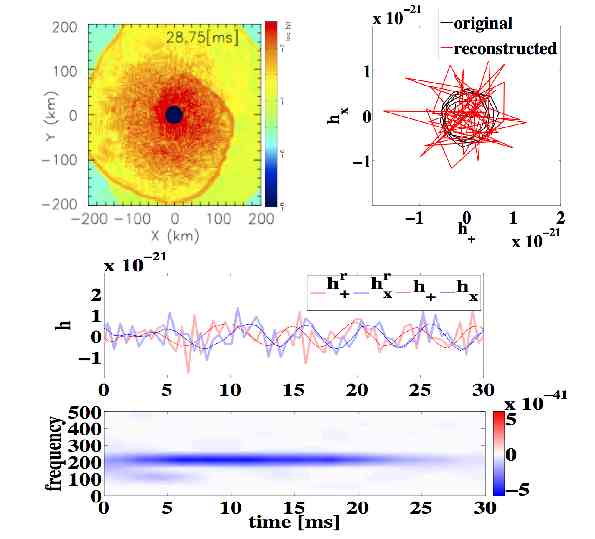}

\vspace{-0.3cm}
\caption{
 The original and reconstructed GWs from one-armed spiral motions in the rapidly rotating supernova core. The top left panel shows the color map of the GW emissivities for the fastest rotating model of \cite{kuroda2014}. The top right and the middle panels present the original and reconstructed $h_+$ and $h_\times$. The bottom panel is the spectrogram for the Stokes V parameter. The observer is assumed to be located at $20$kpc from the source and sitting on the rotation axis of the core.  
}
\label{fig:Vmode_R3p}
\end{center}
\vspace{-25pt}
\end{figure}

\textbf{Simulations and Results {\rm---}} GW waveforms we employ as an input in this letter are taken from 3D general relativistic simulations of rapidly rotating CCSNe by Kuroda {\it{et al.}}~\cite{kuroda2014}. In their numerical models, the authors added by hand almost uniform rotations with periods of $\sim1.6\times 10^{16} \mathrm{cm^2/s}$ at the edge of uniformly rotating core with a size of $R=10^8$ cm (or $M(R)\sim1.3\mathrm{M_\odot}$) to a $\mathrm{15M_\odot}$ non-rotating progenitor model~\cite{WW95} and computed their evolutions up to $\sim 50$ms after core bounce. Since, \cite{2005ApJ...626..350H} and \cite{2000ApJ...544.1016H} reported that the central angular velocity can range
from $\sim0.15$ to $\sim3 \mathrm{rad/s}$ for a non-magnetized $15\mathrm{15M_\odot}$ star at pre-collapse stage,
the most rapidly rotating model with $\pi \mathrm{rad/s}$ is consistent with those stellar evolution calculations. In addition, the specific angular momentum reaches $\sim 1.6\times 10^{16} \mathrm{cm^2/s}$
at the edge of uniformly rotating core with a size of $R=10^8$ cm (or $M(R)\sim1.3\mathrm{15M_\odot}$).
The value is again in good agreement with that of a non-magnetized $15\mathrm{M_\odot}$ model in \cite{2005ApJ...626..350H}.
The GWs were evaluated with the conventional quadrupole formula. They found in their fastest-rotation model that a one-armed spiral motion was formed at $\sim18$ms post-bounce and continued to exist till the end of the simulation (see Fig.~\ref{fig:Vmode_R3p}). 

Using the coherent network analysis, we perform Monte Carlo simulations and following~\cite{hayama2007}. The noise spectrum densities for the four detectors H, L, V, and K are taken from ~\cite{ligodesign,advv,KAGRA}, which the noise used for LIGO, Virgo and KAGR might correspond to the year 2018. The actual locations and orientations of the detectors are adopted in the simulations. A Gaussian, stationary noise 
is produced by first generating four independent realizations of white noise 
with the sampling frequency of $2048$Hz and then passing them 
through the finite impulse response filters having transfer functions that approximately 
match the design curves.
Supernova signals are added to the simulated noises at regular intervals. The location in the sky is assumed to have the right ascension and declination of $(0, 0)$, which is not a special point in Galaxy. In fact, at this sky position and the GPS time of 1045569616, the mean-squared values of the antenna pattern $((F_+^2+F_{\times}^2)/ 2)^{1/2}$ for $h_+$ and $h_\times$ are $0.21$, $0.46$, $0.41$ and $0.49$ for K, H, L and V, respectively. Since the average over the sky is $0.47$, the position considered in this paper is not special indeed. The length of data segments is $100$ms.

Figure~\ref{fig:Vmode_R3p} summarizes the results for the case, in which the GW source is assumed to be located at $20$ kpc from the earth and observed from the rotation axis. A series of short-time ($20$ms) Fourier transforms are calculated to obtain the V parameter from the reconstructed waveforms at different times, to which we refer as the spectrogram here. The interval chosen because the frequencies of the GWs at this early post-bounce phase of post-bounce are demonstrated to be higher than $\sim100\mathrm{Hz}$ by previous researches (e.g. \cite{mueller2013})  We adopt the initial time of each integration as a representative time for the interval. The origin of the time is set to the core bounce.

It is clear from the bottom panel of the figure that right-handed circular polarizations exist indeed at f $\sim200$Hz with a peak amplitude of $\sim5\times10^{-41}$ for the V parameter. The dominance of the right-handed polarization is due to the counter-clockwise rotation of the one-armed spiral pattern (see the top left panel). The signal-to-noise ratio (SNR) is roughly estimated from the so-called radial distance, which corresponds to a network SNR as a detection statistics\cite{hayama2007}: The value of the detection statistics of $0.5$ for the signal should be compared with the one of $0.05$ for the noise in this model. The SNR is then estimated to be $10$, which is, conservatively speaking, significant.

Kuroda et al.~\cite{kuroda2014} argued that the spiral motions are induced by the propagation of acoustic waves Doppler-shifted by the rapid rotation and that the angular velocity of the spiral motion is given by the sum of the angular velocities of the proto\mbox{-}neutron star, $\Omega_{rot}$, and of the acoustic waves, $\Omega_{aco}$. If true, the detection of the circular polarization will provide us with the information on $\Omega_{rot}$,  
since some numerical simulations demonstrated that rotation does not affect $ \Omega_{aco}$ very much and we have almost always $\Omega_{aco} \simeq100$Hz in the early post-bounce phase \citep{mueller2013,kuroda2014}.

\begin{figure}
\begin{center}
\includegraphics[width=0.9\linewidth,height=2.8cm,bb=0 0 950 570]{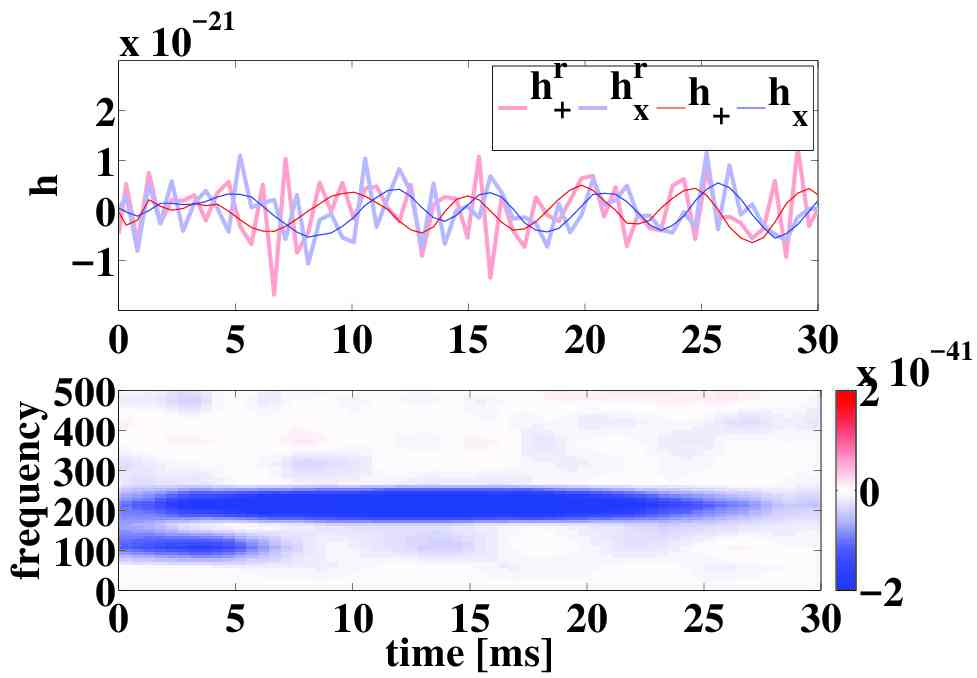}

\vspace{10pt}

\includegraphics[width=0.9\linewidth,height=2.8cm,bb=0 0 950 570]{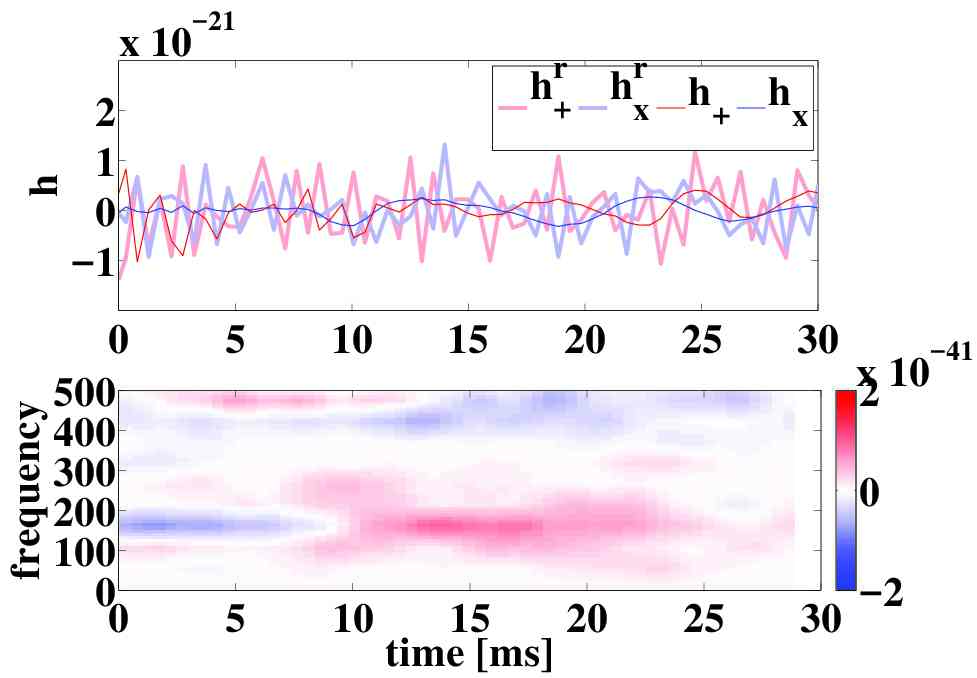}
\vspace{-0.2cm}
\caption{
The spectrograms of the Stokes V parameter for the observers off axis by $45$ (upper panel) and $90$ (lower panel) degrees, respectively. Other parameters are identical to those in Fig.~\ref{fig:Vmode_R3p}.}
\label{fig:Vmode_KTKR345}
\end{center}
\vspace{-30pt}
\end{figure}

The amplitude of the Stokes V parameter depends on the viewing angle. The observer on the rotation axis is certainly the most advantageous. The upper two panels of Figure~\ref{fig:Vmode_KTKR345} display the spectrogram for the GW seen at $45$ degrees from the rotation axis. The circular polarization is clearly seen at $\sim200$Hz also in this case. The SNR is $\sim6$. The magnitude of the V parameter is $-3.9\times10^{-41}$, and the reduction from the previous value, $-5.5\times10^{-41}$, is mainly originated from the cosine of the viewing angle.
For comparison we show the spectrogram of the GW seen from the equator in the lower panel. As expected, the circular polarization disappears almost completely although the root sum square of $+$ and$\times$ modes is almost unchanged from the value $1.0\times10^{-22}\mathrm{[Hz^{-1/2}]}$ for other angles.

\textbf{Discussions \& Conclusion {\rm---}}
We have demonstrated that the circular polarization of the GWs emitted by rapidly rotating supernova cores can be observed up to the opposite end of Galaxy. This is not surprising and can be expected from the back-of-the-envelope
calculations: using the analogy to the GW from binary systems, which is 100\% circularly polarized when observed on the orbital rotation axis, we can evaluate the V-parameter amplitude approximately based on the quadrupole formula as
\begin{eqnarray}
V^{1/2} 
& \sim & 2.1\times 10^{-20} \left(\frac{M_{\rm eff}}{0.5{\rm M_{\odot}}}\right) \left(\frac{R_{\rm eff}}{50{\rm km}}\right)^2 \nonumber \\
&& \times \left(\frac{\omega}{200{\rm Hz}}\right)^2 \left(\frac{D}{10{\rm kpc}}\right)^{-1}.
\end{eqnarray}
Here the observer is assumed to be located at a distance of D from the source on the rotation axis; $M_{\rm eff}$ and $R_{\rm eff}$ are the effective mass and radius of the matter
rotating non-axisymmetrically. Plugging in appropriate values taken from the simulation, we obtain the right order of magnitude. 

We have so far based our investigations of the GW circular polarization on the single model by Kuroda {{\it et al.}}, which may be a concern. This is simply because available calculations are 
quite limited at present. It is true that 3D simulations of CCSN are becoming possible these days, but they are very costly and still in their infancy~\cite{takiwaki2014,tamborra2014}. The number of numerical simulations done
so far is small, particularly for rapidly rotating models. Very recently Nakamura et al.~\cite{nakamura2014} published their 3D Newtonian simulations of the collapse, bounce and explosion, if any, for rapidly rotating cores, employing the so-called light-bulb approximation.  

\begin{figure*}[ht]
\begin{center}

\includegraphics[width=.47\linewidth,height=3cm,bb=0 0 1000 550]{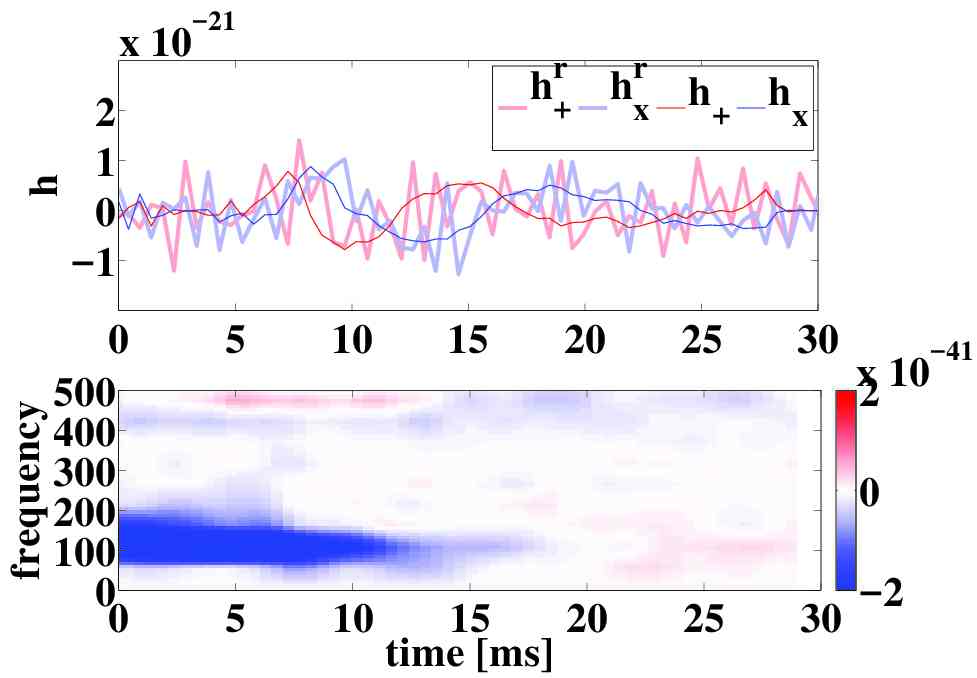}
\includegraphics[width=.47\linewidth,height=3cm,bb=0 0 1000 550]{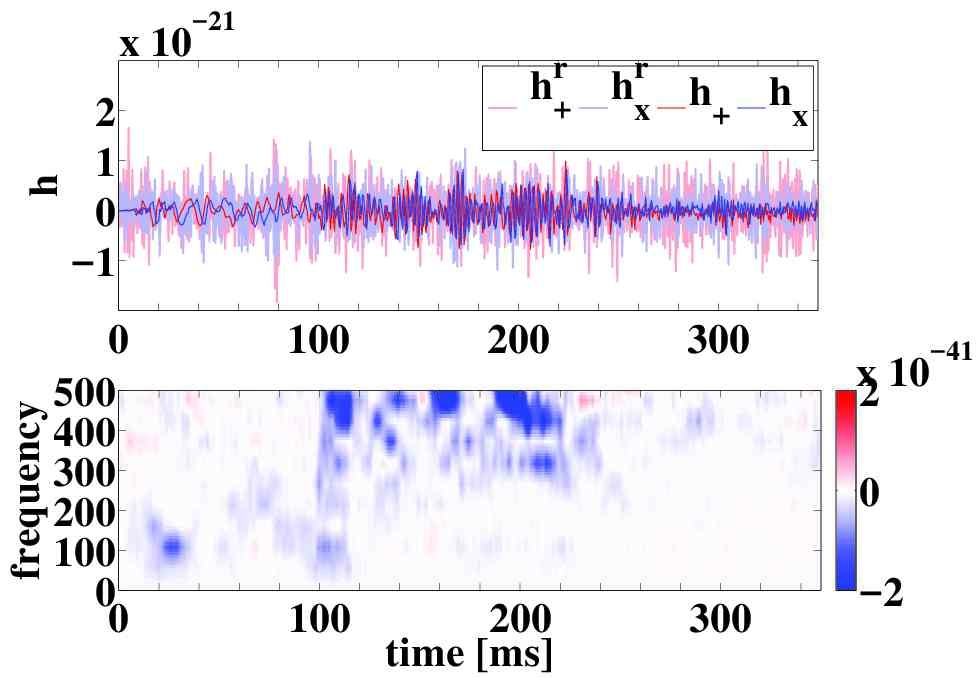}

\vspace{7pt}

\includegraphics[width=.47\linewidth,height=3cm,bb=0 0 1000 550]{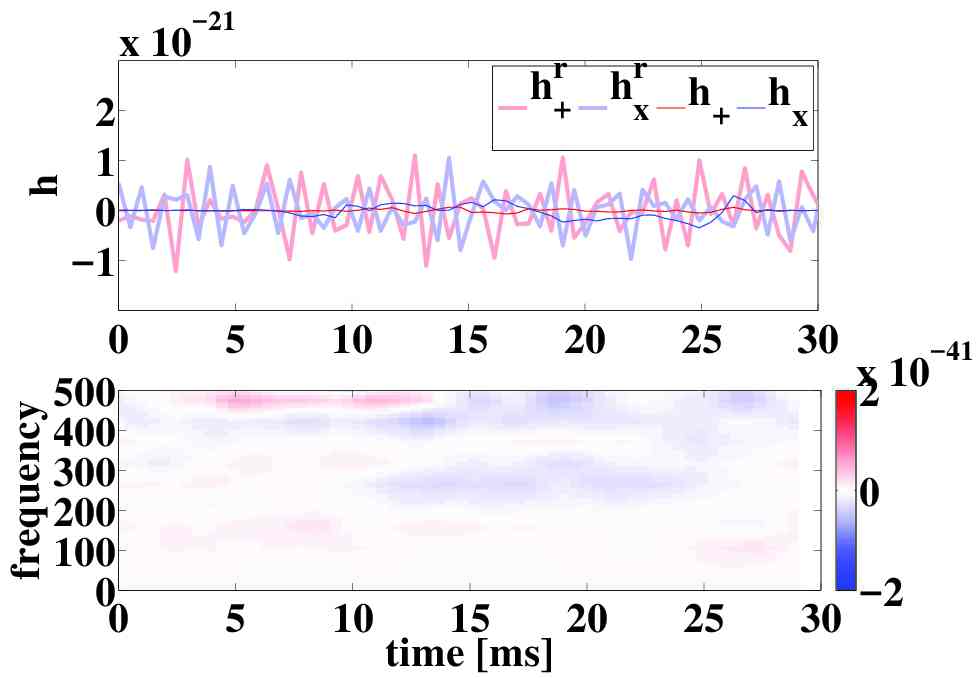}
\includegraphics[width=.47\linewidth,height=3cm,bb=0 0 1000 550]{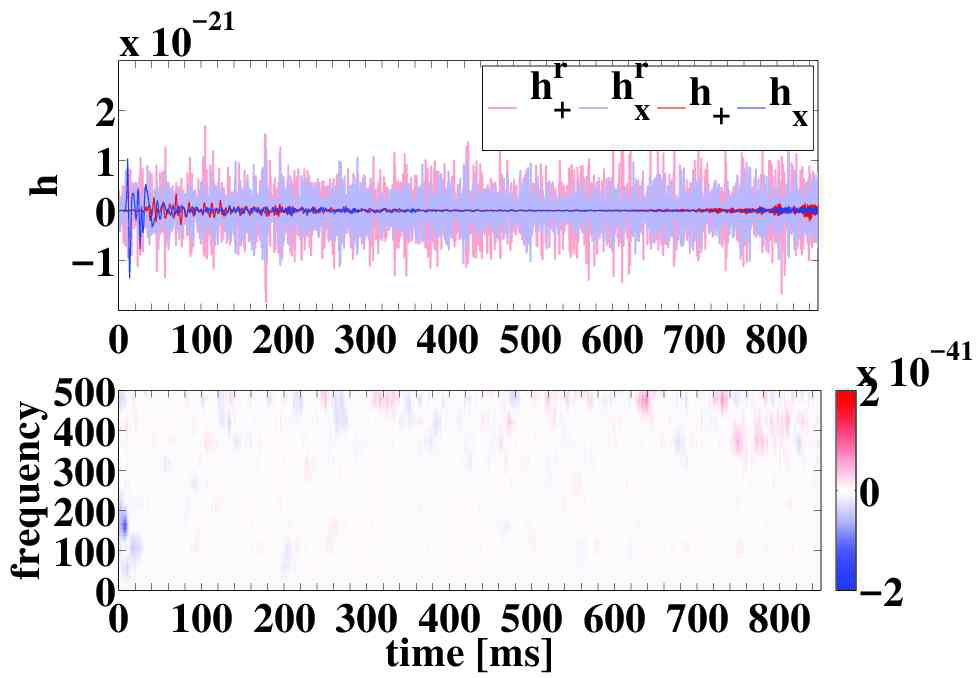}

\vspace{-0.3cm}
\caption{Same as figure~\ref{fig:Vmode_KTKR345}, but for the Kuroda's second fastest rotating (upper two panels in the left column) and non-rotating (lower two panels in the left column) and the corresponding Nakamura's models (the right column). The source is assumed to be located at $D=10\rm{kpc}$ and is observed from the pole.}
\label{fig:Vmode_NAKAMURA2305p}
\end{center}
\vspace{-25pt}
\end{figure*}

We have also calculated circular polarizations for the GW from their rapidly rotating model~\cite{nakamura2014}, which is demonstrated in  Fig.~\ref{fig:Vmode_NAKAMURA2305p}. It is evident from the second-from-the-top plot in the right column that the circular 
polarization appears rather weakly around $30$ms and then reappears more strongly from $100$ms to $200$ms post bounce with short punctuations. 
The former corresponds to what we have discussed so far. As a matter of fact, this model has the same initial rotation velocity as the second fastest model in Kuroda et al.~\cite{kuroda2014}, which indeed produces weak circular polarizations at similar frequencies ($\sim100$Hz). They are hence consistent with each other. What is more remarkable here is the circular polarizations observed at later times. They have higher characteristic frequencies$ \sim500$Hz, possibly due to larger values of $\Omega_{aco}$ at this phase. The stalled shock wave revives and an explosion commences at $\sim240$ms post bounce in this model and the polarization subsides quickly thereafter. 
We estimate that this circular polarization is detectable up to the distance of $\sim10{\rm kpc}$ if it is observed from the rotation axis and this distance will be reduced to $7{\rm kpc}$
if the observer is off axis by $45$ degrees. As expected again, the V parameter is not seen in both Kuroda and Nalamura's non-rotating models.

\begin{acknowledgments}
This work is supported in part by MEXT Leading-edge Research Infrastructure Program,
JSPS Grant-in-Aid for Specially Promoted Research 26000005, the MEXT Grant-in-Aid for the Scientific Research on Innovative Areas "New Developments in Astrophysics Through Multi-Messenger 
Observations of Gravitational Wave Sources" (Nos. 24103005, 24103006) as well as by No. 24244036. KH would like to thank B.~Allen for warm hospitality during his stay in Hannover and S.~D.~Mohanty for valuable comments and encouragements. KH thank M.-K. Fujimoto for continuing encouragements. TK is supported by the European Research Council (ERC; FP7) under ERC Advanced Grant Agreement N$^\circ$ 321263 - FISH
\end{acknowledgments}

\vspace{-20pt}

\bibliography{mybib}

\end{document}